\documentclass[aps,pra,twocolumn,superscriptaddress,showpacs,floatfix]{revtex4}

\usepackage{amssymb,amsmath,graphicx}

\begin{document}

\title{Multi-Manifold Stark Splittings Lift the Rydberg Blockade}

\author{Yurii V. Dumin}
\email[]{dumin@pks.mpg.de,dumin@yahoo.com}
\affiliation{%
Max Planck Institute for the Physics of Complex Systems, \\
Noethnitzer Str.\ 38, 01187 Dresden, Germany}
\affiliation{%
Sternberg Astronomical Institute of Lomonosov Moscow State University, \\
Universitetskii prosp.\ 13, 119992, Moscow, Russia}
\affiliation{Space Research Institute of Russian Academy of Sciences, \\
Profsoyuznaya str.\ 84/32, 117997, Moscow, Russia}

\author{Jan M. Rost}
\email[]{rost@pks.mpg.de}
\affiliation{%
Max Planck Institute for the Physics of Complex Systems, \\
Noethnitzer Str.\ 38, 01187 Dresden, Germany}

\date{November 21, 2016}

\begin{abstract}
The spatial evolution of the Rydberg blockade is studied taking into account
Stark-split energy levels across several manifolds.
We find that the unexpected restoration of a blockaded Rydberg excitation
at small interatomic distances, \textit{e.g.}, experimentally observed by
P.~Schau{\ss}, \textit{et al.} [Nature {\bf 491}, 87 (2012)], can be
explained by the perturbed energy levels from neighboring manifolds that
enter the energy window of excitation defined by the bandwidth of
the exciting laser.
The same mechanism can also explain why the pair correlation
function of Rydberg atoms remains nonzero in the entire region of Rydberg
blockade.
\end{abstract}

%

\maketitle

\section{Introduction}
\label{sec:Intro}

Strong interatomic interaction prevents simultaneous excitation of
more than one Rydberg atom within a compact atomic ensemble by a narrow-band
laser radiation.
This phenomenon of Rydberg blockade was first discussed in
Ref.~\cite{Lukin_01}, and a few years later it was observed
experimentally~\cite{Tong_04,Singer_04}.
Since that time, the Rydberg blockade remains at the heart of quantum
information with ultracold neutral atoms~\cite{Saffman_10}.
Particularly, in the recent years this effect was successfully used
for implementation of the CNOT gate~\cite{Isenhower_10} and demonstration
of entanglement in the two-qubit array of trapped atoms~\cite{Maller_15}.

The concept of Rydberg blockade was involved also in a number of interesting
quantum-optical phenomena, such as a very efficient entanglement of light
with atoms~\cite{Li_13,Weidemueller_13}, coupling a single electron to
a Bose--Einstein condensate~\cite{Balewski_13}, strong effective interaction
between photons~\cite{Peyronel_12,Walker_12} and even creation of the photon
pairs~\cite{Firstenberg_13,Bose_13}, {\it etc.}
Besides, it was suggested that the spontaneous ionization of ultracold
gases, widely studied in the early 2000's~\cite{Gould_01,Bergeson_03},
might be substantially affected by the Rydberg blockade~\cite{Robert_13}.

A detailed spatially-resolved study of the Rydberg blockade, by means of
optical lattices, was performed in Ref.~\cite{Schauss_12}.
These authors measured the probabilities and spatial ordering for
simultaneous excitation of a few atoms (on the time scale
an order of magnitude smaller than the Rabi period).
Particular attention was paid to the case of two atoms.
As a result, it was unexpectedly found that the pair correlation function
exhibits a sharp maximum at small interatomic distances (see Fig.~3a
in the above-cited paper), \textit{i.e.}, the blockade is lifted.
This effect was attributed to imperfection of the detection procedure,
namely, hopping of atoms to the adjacent sites of the optical lattice.
However, as we will demonstrate, the observed behavior can be an inherent
property of the blockade mechanism, which naturally appears if
multi-manifold effects are taken into account.

\section{Theoretical Model}
\label{sec:Model}

Let us consider two Rydberg excited atoms close to each other.
Due to the mutual interaction, the electronic clouds of such atoms
should be deformed with respect to their ionic cores, resulting in
the induced electric dipole moments~$ {\rm \bf d}_e $.
Consequently, the dipolar electric field of each atom will produce
Stark splittings of energy levels in the respective other atom.

When the two atoms approach each other, the energy levels of a manifold,
asymptotically degenerate and resonant with the laser within its bandwidth
at large separations, experience the increasing perturbation and
eventually leave the bandwidth of the exciting irradiation.
Thereby, the Rydberg blockade develops (see Fig.~\ref{fig:Stark_splitting},
which will be discussed in detail below).
However, when the interatomic separation decreases further,
the strongly perturbed energy levels from neighboring Stark manifolds
will enter the excitation band.
Hence, we expect the Rydberg blockade to be lifted at small distances,
about a few Rydberg-atom radii.
(In other words, both atoms can be simultaneously excited.)

To describe this phenomenon, one needs just the standard Van der~Waals
interaction, routinely considered in the physics of ultracold gases.
Indeed, the interaction energy between two induced dipoles
$ {\rm \bf d}_{e1} $ and $ {\rm \bf d}_{e2} $ is
$ E \propto d_{e1} d_{e2} / r^3 $.
However, the absolute values of these dipole moments are not fixed
but rather induced by each other.
Hence, the magnitude of the second dipole is not constant but will be
proportional to the electric field produced by the first dipole,
\textit{i.e.}, $ d_{e2} \! \propto d_{e1} / r^3 $, and \textit{vice versa}.
Therefore, the interaction energy results in
$ E \! \propto d_{e1}^{\,2} / r^6 $.
To study the corresponding splitting of the atomic energy levels,
we shall take into account the terms of the perturbation theory up to
the second order in the electric field \textit{and} to the first order
in the field gradient.

The energy shift for a hydrogen-like Rydberg level~$ n $ then reads
(atomic units are used unless specified otherwise):
\begin{equation}
\delta E_n = E_n +\frac{1}{2n^{2}} =
  g_{1} {\cal E}_z
  - g_{2}{\cal E}_z^2
  + g_{3} \frac{ d {\cal E}_z }{ dz } \, ,
\label{eq:gen_expr_Stark}
\end{equation}
where the $ g_{i} $ ($ g_{2,3} \geqslant 0 $) are
\begin{subequations}
\begin{eqnarray}
g_{1} \!\! & = & \! \frac{3}{2} \, n \Delta \, , \\
g_{2} \!\! & = & \!
  \frac{n^4}{16} \left[ 17 n^2 - 3 \Delta^2  - 9 m^2 + 19 \right] , \\
g_{3} \!\! & = & \! \frac{n^2}{4} \left[ 5 \Delta^2 + 2 n_1 n_2
  + ( n - m ) (m + 1) + 1 \right] .
\end{eqnarray}
\end{subequations}
Here,
$ \cal E $~is the electric field,
$ n $~is the principal quantum number,
$ n_1 $ and $ n_2 $~are the parabolic quantum numbers
($ n_{1,2} \geqslant 0 $),
$ \Delta = n_{1} - n_{2} $, and
$ m $~is the \textit{absolute value} of the magnetic quantum number
(following the Bethe--Salpeter designations~\cite{Bet_Sal});
such that $ n = n_1 + n_2 + m + 1 $.
These quantum numbers satisfy the well-known conditions:
$ m \, \geqslant \, 0 $,\, $n \, \geqslant \, m + 1 $,\, and
$ 0 \, \leqslant \, n_1, n_2 \, \leqslant \, n - m - 1 $.

The $z$-axis is chosen in the direction of the electric field at
the position of the atom experiencing the Stark splitting.
The first two terms in the right-hand side of Eq.~(\ref{eq:gen_expr_Stark})
represent the first- and second-order Stark effect in a uniform
field~\cite{Bet_Sal,Gallagher,Lan_Lif_v3,Yavorsky_80}.
The last term  represents a contribution by the electric-field gradient,
calculated in our recent paper~\cite{Dumin_15}
(the corresponding formula differs somewhat from the early one by
Bekenstein and Krieger~\cite{Bekenstein_70}, which was based on the
WKB approximation).
In principle, it is possible to include here also  higher-order perturbative
terms~\cite{Alliluev_74,Silverstone_78}.
However, these terms are negligible in the present context, as
our subsequent analysis will show%
\footnote{
This is not surprising, because we are interested in the
perturbation of energy levels on the order of the difference between
the states with neighboring values of the principal quantum number
and not on the scale of the binding energy, where the higher-order
corrections should be really significant.
}.

\begin{figure}
\includegraphics[width=5cm]{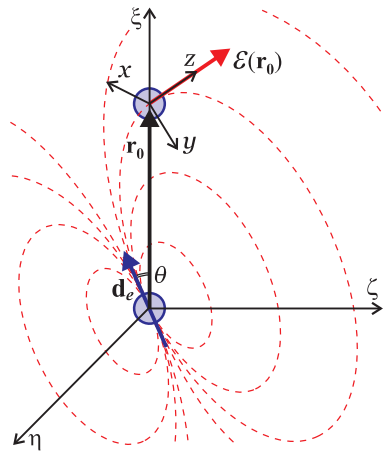}
\caption{Schematic pattern of the electric field
produced in the point~$ {\rm \bf r}_0 $ by
the dipole~$ {\rm \bf d}_e $ located at the coordinate origin.
\label{fig:dipolar_el_field}}
\end{figure}

Next, we calculate the potential
\begin{equation}
\Phi = \frac{ {\bf d}_e \!\! \cdot \! {\bf r}_0 }{ r_0^3 } \,
\label{eq:dipole_potential}
\end{equation}
of the electric dipole~$ {\bf d}_{e} $ at the origin of coordinates
in Fig.~\ref{fig:dipolar_el_field} produced in the point~$ {\bf r}_{0} $,
where the second atom (experiencing the Stark splitting) is located.
The corresponding electric field and its gradient are given by
\begin{subequations}
\begin{equation}
{\cal E}_z = \frac{ d_e }{ r_0^3 } \, ( 1 + 3 \cos^2 \theta )^{1/2}
\label{eq:dipole_field}
\end{equation}
and
\begin{equation}
\frac{ d {\cal E}_z }{ dz } =
  - \frac{ 3 d_e }{ r_0^4 } \:
  \frac{ 3 + 5 \cos^2 \theta }{ 1 + 3 \cos^2 \theta } \, \cos \theta \, ,
\label{eq:dipole_field_grad}
\end{equation}
\end{subequations}
respectively.

We shall consider below in detail the case of two simultaneously-excited
dipoles aligned along~$ {\bf r}_0 $, either in the same direction or
oppositely to each other.
From a physical point of view, the simultaneous excitation corresponds
to processes on a time scale much smaller than the Rabi period,
\textit{e.g.}, as in experiment~\cite{Schauss_12}.
Then,
\begin{equation}
{\cal E}_z = \frac{ 2 d_e }{ r_0^3 }, \qquad
\frac{ d {\cal E}_z }{ dz } = -{\epsilon}_{\theta} \frac{ 6 d_e }{ r_0^4 },
\label{eq:dipole_field_par}
\end{equation}
where
\begin{equation}
{\epsilon}_{\theta} =
\begin{cases}
\hphantom{-} 1, \quad \mbox{at } \theta = 0, \\
            -1, \quad \mbox{at } \theta = {\pi}.
\end{cases}
\label{eq:eps_theta_def}
\end{equation}

Since the electric field is treated here in the classical approximation,
its source is the expectation value of the electric dipole
operator $ \hat{\rm \bf d}_e = - \hat{\rm \bf r}_e $,
where $ {\rm \bf r}_e $~is the radius vector of an electron
inside the atom,
\begin{equation}
d_e = - \langle \, \hat{\xi}_e \, \rangle = - \frac{3}{2} \, n \Delta \, .
\label{eq:dipole_induced}
\end{equation}
This matrix element is well known from the calculation of the first-order
Stark effect~\cite{Bet_Sal,Gallagher,Lan_Lif_v3}.
It is easy to see that
\begin{equation}
{\epsilon}_{\theta} = -{\rm sign}({\Delta}) .
\label{eq:eps_theta_derived}
\end{equation}

In the context of Rydberg blockade, it is convenient to measure
the perturbative energy shift with respect to the energy of the unperturbed
original manifold~$ \bar{n} $ whose blockade is studied
and which is denoted by a bar,
\begin{equation}
\delta E_{\bar{n}} =
  \frac{1}{ 2 {\bar{n}}^2 } - \frac{1}{ 2 n^2 } + \delta E_n\,.
\label{eq:redef_en_shift}
\end{equation}
Furthermore, we scale all lengths and energies with the characteristic size and
energy of the state~$ \bar{n} $.
The scaled quantities will be denoted with a tilde,
\begin{equation}
r_0 = {\bar{n}}^2 \tilde{r}, \quad
E = \tilde{E} / ( 2 {\bar{n}}^2 ).
\label{eq:scaled_Ryd}
\end{equation}
(For conciseness, the scaled radius vector of the atom is written
without subscript~`0'.)

At last, combining~(\ref{eq:gen_expr_Stark}), (\ref{eq:dipole_field_par}),
(\ref{eq:dipole_induced})--(\ref{eq:scaled_Ryd}), we arrive at
the electronic energy-level shifts in the second atom produced by
the first atom:
\begin{eqnarray}
\delta \tilde{E}_{\bar n}^{(2)} & \!\! = \!\! &
1 - \frac{ \bar{n}^2 }{ n^{(2)2} }
  + 9 \bigg[ \frac{1}{{\tilde r}^3} \,
  \frac{ n^{(1)} \, n^{(2)} \, | {\Delta}^{(1)} | \, {\Delta}^{(2)} }%
  { \bar{n}^4 }
\nonumber
\\ &&
  - \, \frac{2}{{\tilde r}^6} \,
  \frac{ g_2^{(2)} n^{(1)2} {\Delta}^{(1)2} }{ \bar{n}^{10} }
  +  \frac{2}{{\tilde r}^4} \,
  \frac{ g_3^{(2)} n^{(1)} {\Delta}^{(1)} }{ \bar{n}^6 } \bigg] .
\label{eq:delta_E(2)_general}
\end{eqnarray}
The number of the particular atom is designated by a superscript in
parentheses (to avoid its confusion with exponents).
The energy shifts in the first atom are given, evidently, by the same
expression with interchanged superscripts.

\begin{figure}
\includegraphics[width=7.8cm]{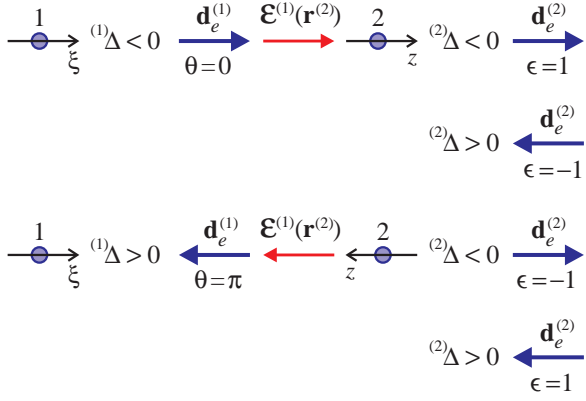}
\caption{Sketch of possible orientations of two dipoles.
\label{fig:two_atoms_interact}}
\end{figure}

Finally, let us consider in detail two particular types of excitations in
our diatomic system:
\begin{eqnarray*}
& \mbox{\bf (a)} &
{| n_1 , n_2 , m \rangle}^{\!(1)}\, {| n_1 , n_2 , m \rangle}^{\!(2)} ,
\; \textit{i.e.},
\\ &&
n \equiv n^{(1)} = n^{(2)} , \quad
\Delta \equiv {\Delta}^{(1)} = {\Delta}^{(2)} ,
\\ &&
g_2^{(1)} \! = g_2^{(2)} , \quad g_3^{(1)} \! = g_3^{(2)} ;
\\
& \mbox{\bf (b)} &
{| n_1 , n_2 , m \rangle}^{\!(1)}\, {| n_2 , n_1 , m \rangle}^{\!(2)} ,
\; \textit{i.e.},
\\ &&
n \equiv n^{(1)} = n^{(2)} , \quad
\Delta \equiv {\Delta}^{(1)} = - {\Delta}^{(2)} ,
\\ &&
g_2^{(1)} \! = g_2^{(2)} , \quad g_3^{(1)} \! = g_3^{(2)} .
\end{eqnarray*}
In other words, both atoms are excited exactly to the same states in case~(a) and
to the states with interchanged parabolic quantum numbers in case~(b).
As is seen in Fig.~\ref{fig:two_atoms_interact}, case~(a) corresponds to
the same orientation of the dipoles (directed to
the right or to the left), while case~(b) features oppositely oriented dipoles
(either towards or away from each other).
We can expect that these two cases represent two limiting situations,
so that the patterns of Rydberg blockade for other orientations will lie
between them.

So, the energy shifts in both atoms under the above-mentioned conditions
will be the same and equal to
\begin{eqnarray}
& \delta \tilde{E}_{\bar n} \! & \! \equiv \,
\delta \tilde{E}_{\bar n}^{(1)} \! = \, \delta \tilde{E}_{\bar n}^{(2)} \! =
  1 - \frac{ \bar{n}^2 }{ n^2 }
  + 9 \bigg[ \frac{1}{{\tilde r}^3} \,
  \frac{ n^2 {\Delta}^2 }{ \bar{n}^4 } \, \epsilon \: {\rm sign} ( \Delta )
\nonumber
\\ &&
  - \, \frac{2}{{\tilde r}^6} \,
  \frac{ g_2 n^2 {\Delta}^2 }{ \bar{n}^{10} }
  +  \frac{2}{{\tilde r}^4} \,
  \frac{ g_3 n \Delta }{ \bar{n}^6 } \bigg] ,
\label{eq:delta_E_symmet}
\end{eqnarray}
where
\begin{equation}
{\epsilon} =
\begin{cases}
\hphantom{-} 1, \; \mbox{for parallel dipoles}, \\
            -1, \; \mbox{for anti-parallel dipoles}.
\end{cases}
\label{eq:eps_def}
\end{equation}
Note that the first and second terms in square brackets in
Eq.~(\ref{eq:delta_E_symmet}) result from the first- and second-order Stark
effect in the uniform field, respectively; while the third term comes from
the first-order perturbation by the electric field gradient.

\begin{figure}
\includegraphics[width=8.2cm]{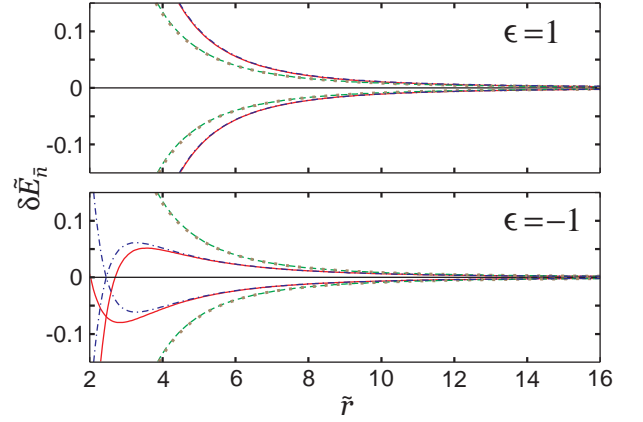}
\caption{
Energy level shifts~$ \delta \tilde{E}_{\bar{n}} $ for
$ n = \! \bar{n} \! = 43 $; $ m = 0 $; $ n_1 \! = \! 0 $ and 42
(the most disturbed sublevels) taking into account
only the first-order uniform-field Stark effect (dashed green curves),
the first- and second-order uniform-field effects (dotted brown curves),
the first-order uniform-field and gradient-term effects
(dot-and-dashed blue curves), and all three contributions
in~(\ref{eq:delta_E_symmet}) (red solid curves).
\label{fig:Relative_contrib}}
\end{figure}

\begin{figure*}
\includegraphics[width=16cm]{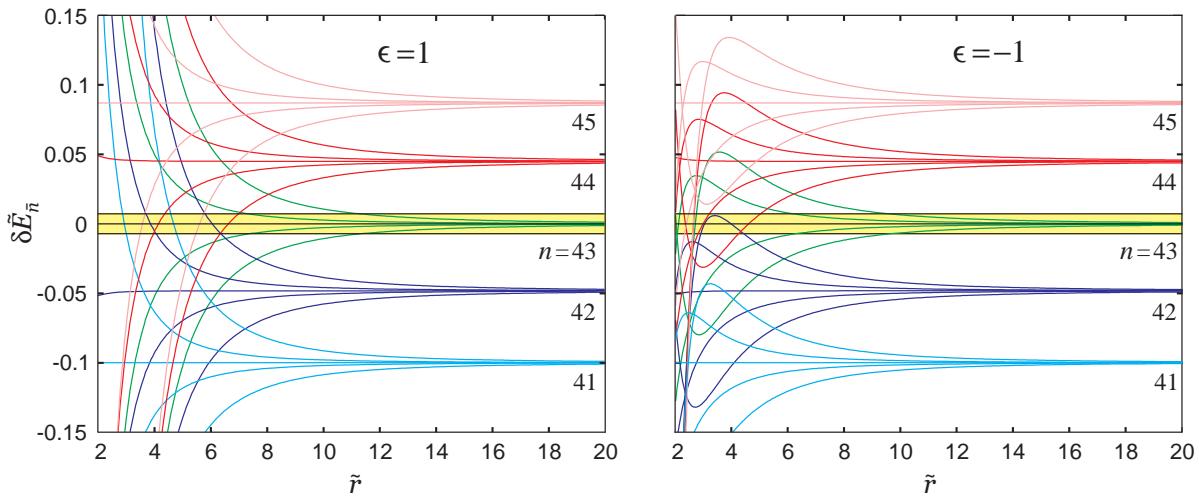}
\caption{Stark splitting~$ \delta \tilde{E}_{\bar{n}} $ of energy levels
as function of the interatomic distance~$ \tilde{r} $ for
$ n = \! \bar{n} \! = 43, \: n_1 \! = \! 0, 10, 21, 32, 42 $ (green curves),
$ n \! = \! 42, \: n_1 \! = \! 0, 10, 20, 31, 41 $ (blue),
$ n \! = \! 41, \: n_1 \! = \! 0, 10, 20, 30, 40 $ (cyan),
$ n \! = \! 44, \: n_1 \! = \! 0, 11, 22, 32, 43 $ (red), and
$ n \! = \! 45, \: n_1 \! = \! 0, 11, 22, 33, 44 $ (magenta).
The horizontal shaded (yellow) strip denotes the energy excitation band
of laser irradiation, $ \Delta \tilde{E} = 1.44{\cdot}10^{-2} $.
\label{fig:Stark_splitting}}
\end{figure*}

To reveal effects of the different perturbative terms, we have drawn in
Fig.~\ref{fig:Relative_contrib} a series of curves taking into account
the combinations of them.
For parallel dipoles~($ \epsilon = 1 $) the following conclusions can be
drawn:
\begin{itemize}
\item
The dotted (brown) curves are almost indistinguishable from
the dashed (green) curves, implying that the second-order Stark shift
is negligible compared to the first-order Stark shift in the uniform field.
\item
There is an appreciable yet not qualitative difference between
the dashed (green) and dot-dashed (blue) curves.
Consequently, the gradient-term Stark effect is noticeable.
\item
Since the solid (red) curves almost coincide with the dot-dashed (blue)
curves, the second-order Stark effect is again negligible.
\end{itemize}

With regard to anti-parallel dipoles~($ \epsilon = -1 $)
the situation is as follows:
\begin{itemize}
\item
Again, the dotted (brown) curves are indistinguishable from
the dashed (green) curves. Hence, the second-order Stark shift can be
ignored if the gradient term is not taken into consideration.
\item
The difference between the dashed (green) curves and the dot-dashed
(blue) curves is larger than in the case~$ \epsilon = 1 $ and becomes even
qualitative (the corresponding curves are bent in the opposite direction
at small distances). Therefore, the gradient-term Stark effect plays
an important role.
\item
One also sees that the solid (red) curves are shifted downwards with
respect to the dot-dashed (blue) curves at small~$ \tilde{r} $.
Hence, in principle, the second-order Stark effect becomes noticeable but
does not play a crucial role.
\end{itemize}

In summary, we can say that the gradient term is always important and can
even qualitatively change the behavior of energy curves at small~$ \tilde{r} $
in the case of anti-parallel dipoles.
On the other hand, one can usually neglect the second-order Stark effect in
the uniform field.
It becomes noticeable only at small distances in the case of anti-parallel
dipoles, when the first-order perturbations by the uniform field and
the gradient term compensate each other to a large extent.

Returning to the problem of breaking the Rydberg blockade, we show in
Fig.~\ref{fig:Stark_splitting} examples of a few Stark-split energy levels
for the parameters of experiment~\cite{Schauss_12}, where~$ {}^{87}{\rm Rb} $
with $ \bar{n} = 43 $ and $ m = 0 $ was used.
For better visibility, the energy excitation band~$ \Delta \tilde{E} $
(yellow stripe) was taken here 100~times larger than in the experiment.
Besides, to avoid excessive complication of the figure, we have not shown
the avoided crossings, because they do not affect
the \textit{total density of energy levels} in the excitation band.

The pattern of levels for~$ \epsilon = 1 $ is approximately symmetric
with respect to their asymptotic degenerate positions,
in qualitative agreement with our earlier considerations~\cite{Dumin_14}.
On the other hand, for anti-parallel dipoles ($ \epsilon = -1 $)
the levels show nonmonotonic and asymmetric behavior.

At sufficiently large distances, all levels of the $ n=43 $~manifold are
within the excitation band.
When $ \tilde{r} $~decreases, these levels begin to deviate from
the horizontal line and leave the excitation band such that the Rydberg
blockade develops.
However, when the distance decreases further, down to $ \tilde{r} = 5-6 $,
the strongly-perturbed energy levels from the neighboring Stark manifolds
(or even the levels from the same manifold~$ n=43 $ bent in the opposite
direction) enter the excitation band, thereby lifting the Rydberg blockade.
As can be seen in this figure, at~$ \epsilon = 1 $ the blockade is lifted by
both the upper- and lower-lying Stark manifolds.
In contrast, for~$ \epsilon = -1 $ the contribution comes mostly from
the upper-lying manifolds.

\begin{figure*}
\includegraphics[width=16.5cm]{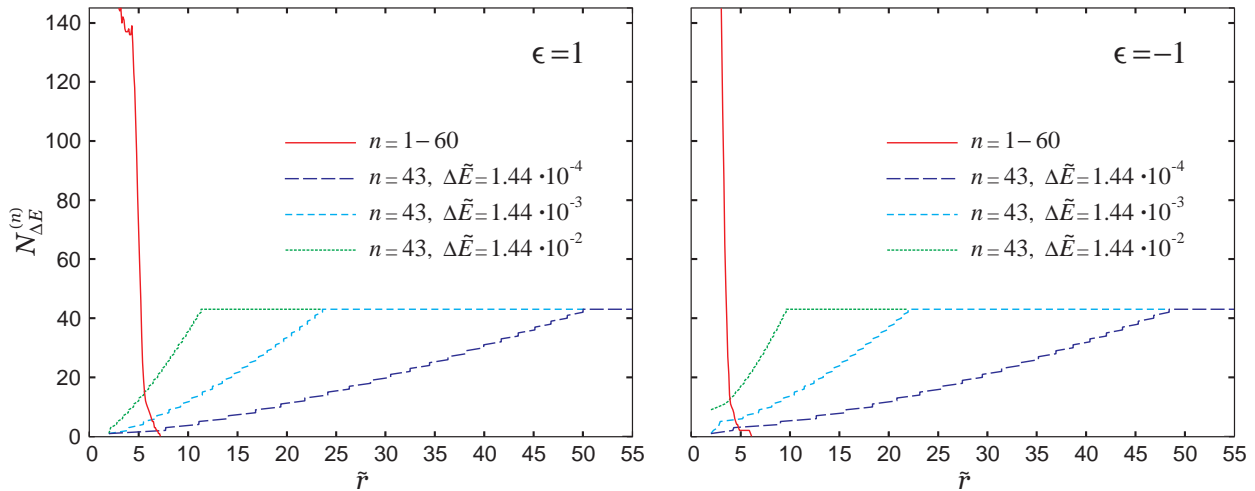}
\caption{The number of energy levels~$ N_{\Delta E}^{(n)} $
within the running window~$ \Delta \tilde{r} = 1 $ for various
bandwidths~$ \Delta \tilde{E} $ of the exciting irradiation
and different Stark manifolds~$ n $.
\label{fig:Number_of_levels}}
\end{figure*}

The total effect of the level shifts is illustrated in
Fig.~\ref{fig:Number_of_levels}, which represents the total number of
energy levels~$ N_{\Delta E}^{(n)} $ in the excitation
band~$ \Delta \tilde{E} $ within a running
window~$ \Delta \tilde{r} = 1 $ (\textit{i.e.}, the characteristic size
of the Rydberg atom).
The dashed and dotted (blue, cyan, and green) curves show the number
of levels of the basic manifold~$ n=43 $ remaining within the excitation
band as function of the distance.
Not unexpected, the narrower the excitation band, the larger is
the Rydberg blockade zone.
The experimental bandwidth in~\cite{Schauss_12} was very small,
$ \Delta \tilde{E} = 1.44{\cdot}10^{-4} $, corresponding to
the long-dashed (blue) curves in Fig.~\ref{fig:Number_of_levels} that drop
off at the largest distance ($ \tilde{r}{\leqslant}50 $) and vary most
gradually.

The red curve, sharply rising at small distances indicates lifting
of the Rydberg blockade by the Stark-split energy levels from
the neighbouring manifolds.
It is approximately independent of~$ \Delta \tilde{E} $, since
the strongly-perturbed energy levels enter the excitation band
almost vertically.
The upper limit of~$ n $, until which the energy levels are counted,
is somewhat arbitrary but certainly limited by the range of validity
of the perturbative Stark splitting.
The latter is roughly given by the condition that the energy shift
should be less than the absolute value of the unperturbed energy:
$ | E_{\bar{n}} | - | E_n | \lesssim | E_n | $, implying
$ n \lesssim \sqrt{2} \, \bar{n} $.
Hence, for $ \bar{n} = 43 $, we get an upper limit of $ n \approx 60 $.
On the other hand, it can be easily seen that perturbation theory
is applicable to all the low-lying levels, because the criterion
$ | E_n | - | E_{\bar{n}} | \lesssim \, | E_n | $ is always satisfied
for $ n < \bar{n} $.

Note that despite the qualitatively different patterns of individual
energy levels for~$ \epsilon = 1 $ and~$ -1 $
(Fig.~\ref{fig:Stark_splitting}), the resulting effect on the Rydberg
blockade is almost the same: the left and right panels of
Fig.~\ref{fig:Number_of_levels} are very similar to each other;
only the blockade-lifting red peak is narrower for the anti-parallel
dipoles.

The values for the outer boundary of the Rydberg blockade zone
($ \tilde{r} = 50 $ at $ \Delta \tilde{E} = 1.44{\cdot}10^{-4} $)
and the inner boundary, where the blockade breaks down ($ \tilde{r} = 5 $),
correspond very well to the measurements presented in~Fig.~3a
of~\cite{Schauss_12}.
For the comparison of the experimental and theoretical plots,
one should keep in mind that 1~$\mu$m corresponds to 9.8~dimensionless
units of length used in the present paper.
In passing, we note that the additional excitation zones, formed by
the neighboring Stark manifolds, merge into one broad shell about
$ \tilde{r} = 0 $.
Therefore, the individual thin shells (qualitatively discussed
in~\cite{Dumin_14}) can be hardly resolved under realistic experimental
conditions.

\section{Discussion}
\label{sec:Discus}

Strictly speaking, our study refers only to hydrogen-like atoms without
quantum defects.
They were taken into account recently in~\cite{Derevianko_15},
whose authors employed the idea of quasi-molecular levels of two nearby
Rydberg atoms%
\footnote{
The formation of long-range molecules from two
Rydberg atoms was discussed over a decade ago for both the case with
substantial quantum defects~\cite{Boisseau_02} and for hydrogen-like
atoms~\cite{Flannery_05}.
}.
Calculating in detail such levels for the 100s~state of rubidium, they
obtained the probabilities of detrimental excitation (or the ``molecular
loss rates'') for a few unblocked shells.
It was found that the most significant effect,
\textit{i.e.}, the largest excitation rate, is produced by the outermost
unblocked shell.
Since it is affected least by the quantum defects, neglecting them
should be reasonable in many situations.

Relevant is also the size of the outermost shell~$ R_{\rm b} $,
which we estimate in the following to exponential accuracy
(\textit{i.e.}, ignoring pre-exponential factors on the order of unity).
In the framework of the quasi-molecular~(qm) model one gets
\begin{subequations}
\label{eq:Rb}
\begin{equation}
R_{\rm b}^{\rm (qm)} \! \propto n^{8/3} \approx n^{2.67} ,
\label{eq:R_b_quasi-mol}
\end{equation}
see Sec.~II in Ref.~\cite{Derevianko_15}.
On the other hand, our earlier treatment of the Rydberg blockade in terms
of the Stark effect in the uniform field~(uS) gives
\begin{equation}
R_{\rm b}^{\rm (uS)} \! \propto n^{7/3} \approx n^{2.33} ,
\label{eq:R_b_unif-Stark}
\end{equation}
see formula~(14) in Ref.~\cite{Dumin_14}.
At last, the same treatment for the strongly nonuniform field~(nS)
results in formula~(36) of Ref.~\cite{Dumin_15} with
\begin{equation}
R_{\rm b}^{\rm (nS)} \! \propto n^{9/4} \approx n^{2.25} .
\label{eq:R_b_nonunif-Stark}
\end{equation}
\end{subequations}
Hence, all three approximations give  similar results with
the quasi-molecular model predicting the steepest dependence on
the principal quantum number.

Clearly, the quasi-molecular treatment appears to be more
accurate, because it takes into account the structure of energy levels
of a particular atom at the expense of a cumbersome computation and
the loss of general relevance%
\footnote{
Several features of the Rydberg blockade specific to rubidium atoms were
studied in~\cite{Reinhard_07}, but the possibility that the blockade
can be broken at some radii was not taken into consideration.
}.
Our approach based on Stark splitting is of complementary value, offering
a simple and straightforward determination of the blockade breaking from
a general perspective, omitting possible additional features, different
for each atom.
It can be especially valuable for reasonable estimates in situations
where it is unknown in advance which chemical element and which of its
states could be employed best for the quantum information processing.

Besides and most importantly, our approach explains transparently
why the pair correlation function of Rydberg atoms is never equal to
zero in the blockaded volume (\textit{e.g.}, Fig.~3a of
Ref.~\cite{Schauss_12}), whereas this fact is not so clear from
the quasi-molecular model~\cite{Derevianko_15}.

Yet another method of treating the interactions between Rydberg atoms
is the multipole expansion~\cite{Flannery_05}.
It is based on the same physical assumptions as our calculation
of the gradient-term Stark effect~\cite{Dumin_15}, namely, the first-order
perturbation theory with respect to the product state of two unperturbed
atoms.
However, the multipole expansion results in a lot of cumbersome terms,
making a subsequent analysis necessary which of them are really
important in the respective context.
In contrast, the treatment based on the gradient term ``automatically''
takes into account the most important contributions just from the beginning.
Hence, it has the same physical accuracy as the multipole techniques but
works more efficiently for our needs.

Finally, let us mention that also \textit{non-additive} perturbations in
the systems containing more than two atoms can break the Rydberg blockade.
Such a situation for three atoms was considered in paper~\cite{Pohl_09}.

\section{Conclusions}
\label{sec:Concl}

In summary, we have developed an analytical model, sufficiently general to
be used in order to refine certain properties of the Rydberg blockade.
Firstly, we conclude that lifting the Rydberg blockade at small
interparticle distances has a universal physical reason, rather than
being a result of experimental imperfections.
This effect should be kept in mind in the design of future
high-precision experiments, \textit{e.g.}, for quantum information
processing.

Secondly, our description in terms of the Stark-split manifolds naturally
explains also why the probability of excitation remains nonzero in
the major part of the blockaded volume.
This is yet another inherent property of the Rydberg blockade, rather
than the effect of ``imperfect removal of the ground-state atoms''
in~\cite{Schauss_12}.

Finally, the treatment of the Rydberg blockade performed in the present paper
refers to the coherent (simultaneous) excitation of Rydberg atoms.
If the excitation of atoms is incoherent (sequential), as assumed in
our previous works~\cite{Dumin_14, Dumin_15}, the Rydberg blockade should
also be lifted at small distances but can possess some specific features,
which still have to be studied in detail.
This may be interesting, \textit{e.g.}, for the experiments with
ultracold plasmas performed on much longer time scales~\cite{Robert_13}.

\acknowledgments

One of the authors (YVD) is grateful to
R.~C{\^o}t{\'e} and
P.~Schau{\ss}
for valuable discussions and advise.


\end{document}